# Power and Bit Allocation for Wireless OFDM Channels with Finite-Rate Feedback and Subcarrier Clustering

**Wiroonsak Santipach\*, Kritsada Mamat, Ake Tonsirisittikun and Kaemmatat Jiravanstit**

---

## ABSTRACT

The study investigated the allocation of transmission power and bits for a point-to-point orthogonal frequency-division multiplexing channel assuming perfect channel information at the receiver, but imperfect channel information at the transmitter. Channel information was quantized at the receiver and was sent back to the transmitter via a finite-rate feedback channel. Based on limited feedback from the receiver, the corresponding transmitter adapted the power level and/or modulation across subcarriers. To reduce the amount of feedback, subcarriers were partitioned into different clusters and an on/off threshold-based power allocation was applied to subcarrier clusters. In addition, two options were proposed to interpolate a channel frequency response from a set of quantized channel gains and apply the optimal water-filling allocation or a greedy bit allocation based on channel interpolation. Proposed schemes with finite feedback rates were shown to perform close to the optimal allocation without a feedback-rate constraint. In the numerical example, channel capacity decreased about 6% from the optimum when one bit of feedback per subcarrier was used.

## INTRODUCTION

Orthogonal frequency-division multiplexing (OFDM) has been widely used in current wireless communication systems; for example, digital audio broadcast, digital video broadcast, IEEE Std. 802.11 (WiFi), IEEE Std. 802.16 (WiMAX) and Long-Term Evolution (Institute of Electrical and Electronics Engineers Computer Society, 2004), due to its high spectral efficiency and robustness to frequency-selective fading channels. In OFDM, the data stream is divided and transmitted over many low rate parallel subchannels, which help to increase symbol duration and reduce inter-symbol interference (ISI) (Weinstein and Ebert, 1971; Cimini, 1985). In other words, OFDM transforms a frequency-selective channel into a collection of parallel flat-fading channels.

The performance of an OFDM system depends on the transmission power, the transmission bandwidth and also the channel information available at both a transmitter and receiver. With channel information at a receiver, coherent detection can be performed while a transmitter with channel information can adapt modulation and power allocation across subchannels or subcarriers (Liu *et al.,* 2009). A receiver can

---

Department of Electrical Engineering, Faculty of Engineering, Kasetsart University, Bangkok 10900, Thailand.
\*   Corresponding author, email: wiroonsak.s@ku.ac.th

estimate a channel from the pilot signal and several pilot-aided channel estimation schemes have been proposed by Tong *et al.* (2004) and various references therein. However, a transmitter may not be able to estimate a forward channel by itself (for example, in a frequency-division duplex) and has to rely on the receiver for a channel-information update via a finite-rate feedback channel.

Based on the channel information, the receiver can optimize the tranmit power over all subcarriers so that it maximizes the channel capacity and then can feed back a set of optimized power levels to the transmitter. To reduce the feedback rate, an on/off threshold-based power allocation was proposed and was shown to perform close to the optimal water-filling allocation (Bingham, 1990; Jang and Lee, 2003; Tse and Visawanath, 2005; Sun and Honig, 2008). Channel update at the transmitter can also be used to adapt modulation (Wong *et al.,* 1999; Karami *et al.,* 2010). The current study proposed to further reduce the feedback by partitioning adjacent subcarriers into different clusters and subsequently applying the on/off power allocation. Thus, the feedback rate to relay power levels is reduced to essentially one bit per cluster. The numerical results from the current study have shown that selecting the appropriate cluster size and threshold can boost the performance of the proposed scheme close to the optimum, but with much smaller feedback. Santipach (2010) has considered the on/off allocation with imperfect channel information at both the transmitter and receiver, but without subcarrier clustering.

Instead of a set of power levels, the receiver can relay channel estimates back to the transmitter, which then can allocate transmit power based on the channel estimates. Since feedback is limited, the current study proposed to quantize the channel gain of only one subcarrier in a cluster. Based on a set of quantized channel gains for some specific subcarriers, the transmitter reconstructs the entire channel frequency response by interpolating the rest of the channel gains, and finds the water-filling solution based on channel interpolation. Liu *et al.* (2009) proposed a similar scheme; however, the set of channel gains that was relayed from the receiver was assumed to be perfect.

In addition to the transmission power levels, the current study proposed to optimize the bit allocation across subcarriers based on either linear or quadratic channel interpolation. For a given data rate and probability of error, an iterative bit allocation method proposed by Wong *et al.* (1999) was applied that minimizes the total transmission power. As a result, each subcarrier may use a different modulation scheme depending on the estimate of the channel condition.

## MATERIALS AND METHODS

### Channel model

We consider a point-to-point discrete-time OFDM channel with $N$ subcarriers. An $M$-tap channel impulse response is denoted by an $M \times 1$ vector

$$\mathbf{h} = \begin{bmatrix} h_0 & h_1 & \cdots h_{M-1} \end{bmatrix}^T. \tag{1}$$

Where $h_m$ is the $m^{\text{th}}$ channel tap and $[\cdot]^{\text{T}}$ is the transpose operation.

Assuming a rich scattering and non line-of-sight environment, the channel is hence Rayleigh faded. We also assume that all channel paths are approximately equal and thus, have equal power. For the Rayleigh fading and uniform power delay profile, each channel tap $h_m$ is modeled by an independent complex Gaussian random variable with zero mean and variance $\sigma_m^2 = \dfrac{1}{M}$. Thus, $\sum_{m=0}^{M-1}\sigma_m^2 = 1$. A cyclic prefix for each OFDM symbol is assumed to be long enough to suppress ISI. Applying a discrete Fourier transform (DFT) gives a channel frequency response of the $i$th subcarrier as shown in Equation 2:

$$H(i) = \sum_{m=0}^{M-1} h_m \mathrm{e}^{-\frac{j2\pi mi}{N}}, \quad 0 \leq i \leq N-1. \tag{2}$$

A frequency-selective channel is converted to $N$ parallel flat-fading subchannels and the output signal of the DFT at the receiver can be written as Equation 3:

$$y(i) = \sqrt{P_i} H(i)x(i) + n(i), \quad 0 \le i \le N-1, \tag{3}$$

where $x(i)$ is a transmitted symbol on the $i$th subchannel, $P_i$ is a transmit power allocated for the $i$th subchannel, and $n(i)$ is an additive white Gaussian noise (AWGN) with zero mean and variance $\sigma_n^2$. A corresponding sum capacity over all subchannels is given by Equation 4:

$$C = \sum_{i=0}^{N-1} E\left[ \log(1 + \frac{P_i |H(i)|^2}{\sigma_n^2}) \right] \tag{4}$$

where the expectation is over a joint distribution of $H(i)$ and there exists a total-power constraint given by $\sum_{i=0}^{N-1} P_i \le P_T$. Given the set of channel responses $\{H(0), H(1),\ldots, H(N-1)\}$, the optimal set of transmit power levels $\{P_0, P_1,\ldots,P_{N-1}\}$ that maximizes the sum capacity is obtained by the well-known water-filling solution (Bingham, 1990; Tse and Viswanath, 2005).

The sum capacity bit error rate (BER) is another important performance metric for a communication channel. We assume that the transmitter can also adapt a modulation scheme on all subcarriers based on the channel frequency response. The BER for the $i$th subcarrier is lower bounded by Equation 5:

$$\text{BER}_i \ge \frac{P_{e,i}}{\log_2 M_i} \tag{5}$$

where $P_{e,i}$ denotes the probability of symbol error for the $i$th subcarrier and $M_i$ denotes the number of different symbols transmitted on the $i$th subcarrier. Assuming that the data stream on each subcarrier is transmitted by $M_i$-ary quadrature amplitude modulation (QAM) with a square signal constellation (4-QAM, 16-QAM, or 64-QAM), probability of symbol error can be accurately approximated by Equation 6:

$$P_{e,i} = E\left[ 4Q\left( \sqrt{\frac{3P_i |H(i)|^2}{\sigma_n^2 2^{c_i - 1}}} \right) \right] \tag{6}$$

where Q function is defined by Equation 7:

$$Q(x) = \frac{1}{\sqrt{2\pi}} \int_x^\infty e^{-t^2/2} dt , \tag{7}$$

where $c_i = \log_2 M_i$ is the number of bits per symbol transmitted on the $i$th subcarrier, and the expectation is over the distribution of $|H(i)|^2$. A system BER is defined to be an average BER over $N$ subcarriers by Equation 8:

$$\text{BER} = \frac{\sum_{i=0}^{N-1} \text{BER}_i}{N} \tag{8}$$

With channel information, the transmitter can calculate the optimal $\{C_i\}$ that minimizes the total transmit power for a given total data rate and the BER.

At the receiver, channel information can be estimated from the pilot signal. Thus, the receiver can obtain the optimal set of transmit power or modulation schemes based on the channel estimation, then the receiver feeds the set back to the transmitter. Here we assume that the receiver can estimate the channel perfectly. Since the feedback rate is limited, the optimal power and bit allocation need to be quantized. Hence, the associated sum capacity and system BER depend on the quantization error, which in turn, depends on the feedback rate. Given a feedback rate, we propose power and bit allocation schemes in the following sections.

## On/off power allocation with subchannel clustering

To reduce the number of bits to quantize the set of transmitted power, we consider an on/off power allocation in which an equal power is allocated for a subcarrier if its channel gain exceeds certain thresholds, and zero power is allocated otherwise (Sun and Honig, 2008). The on/off power allocation is clearly suboptimal, but performs close to the optimum in a large signal-to-noise ratio regime. The number of bits required to relay the transmitted power is only 1 bit per subcarrier. To further reduce the number of bits, we exploit the correlation among nearby subcarriers. Since the number of channel taps is much lower than the total number of subcarriers, channel gains of adjacent subcarriers are highly correlated. Therefore, we group subchannels into clusters, which consist of $R$ neighboring subcarriers (except possibly the last cluster). The number of total clusters is denoted by $K = \lceil N/R \rceil$ where $\lceil \cdot \rceil$ is the ceiling function. The average channel gain squared over the $k^{\text{th}}$ cluster, where $0 \leq k \leq K\text{-}2$, is given by Equation 9:

$$\frac{1}{R} \sum_{r=0}^{R-1} |H(Rk + r)|^2 \qquad (9)$$

and the average of the last cluster is given by Equation 10:

$$\frac{1}{N-(K-1)R} \sum_{i=(K-1)R}^{N-1} |H(i)|^2 . \qquad (10)$$

We propose to allocate equal power for all subcarriers in the cluster whose average channel gain squared exceeds threshold $\mu$ and to allocate zero power otherwise. Thus, the number of subcarriers with nonzero power or activated subcarriers is given by Equation 11:

$$N_A = \sum_{k=0}^{K-2} 1_\mu \left( \frac{1}{R} \sum_{r=0}^{R-1} |H(Rk + r)|^2 \right) +$$

$$1_\mu \left( \frac{1}{N-(K-1)R} \sum_{i=(K-1)R}^{N-1} |H(i)|^2 \right) \qquad (11)$$

where an indicator function is provided by Equation 12:

$$1_\mu(x) = \begin{cases} 1: & x \geq \mu \\ 0: & x < \mu \end{cases} \qquad (12)$$

and the transmitted power allocated for each activated subcarrier is $P_T/N_A$. The corresponding sum capacity will depend on the number of subcarriers in a cluster, $R$, and the threshold $\mu$. Feeding back the set of transmit power levels, the proposed on/off power with subchannel clustering requires 1 bit per cluster with total feedback bits equal to $K + \log_2 R$ bits, which can be significantly less than $N$. We note that for $K = N$, the proposed scheme reverts back to the conventional on/off power without clustering.

## Water-filling power allocation with channel interpolation

Instead of a set of transmitted power levels for subcarriers, the receiver in this scheme feeds back a set of channel gains of certain subcarriers to the transmitter. From the given set of channel gains, the transmitter then interpolates the rest of the channel response and allocates subcarrier powers by the water-filling solution based on the channel interpolation.

With perfect channel information, the receiver quantizes the squared channel gains of the $K$ subcarriers, which are $R$ subcarriers apart (except possibly the last pair). We denote the quantized squared channel gain for the $i^{\text{th}}$ subcarrier by $\alpha(i) = Q_s (| H(i) |_2)$, where $Q_s (\cdot)$ is a uniform scalar quantizer. Thus, with $B$ bits, the receiver feeds back the following set of $K$ quantized squared channel gains $\{\alpha(0), \alpha(R), \alpha(2R),\ldots, \alpha((K-2)R), \alpha(N)\}$.

### Linear interpolation

At the transmitter, the rest of the squared channel gains can be interpolated from the set of $K$ quantized squared channel gains. First, we consider a linear interpolation. For the first $K-1$ clusters, the interpolated squared channel gain is given by Equation 13:

$$\hat{\alpha}(Rk + r) = \alpha(Rk) + (\alpha(R(k+1)) - \alpha(Rk)) \frac{r}{R} \quad (13)$$

where $1 \leq r \leq R$ and $0 \leq k \leq K - 2$. For the last cluster, Equation 14 applies:

$$\hat{\alpha}(R(K-1) + r) = \alpha(R(K-1)) + (\alpha(N) -$$

$$\alpha(R(K-1))) \frac{r}{N - R(K-1)} \qquad (14)$$

where $1 \leq r \leq N - R(K-1)$.

With the set of interpolated squared channel gains $\{\hat{\alpha}(i)\}$, the water-filling solutions for transmission power can be found by solving

the Equations 15 and 16:

$$P_i = \left[ \gamma - \frac{\sigma_n^2}{\hat{\alpha}(i)} \right]^+ \qquad (15)$$

for $0 \le i \le N-1$, where the function

$$[x]^+ = \begin{cases} x : & x \ge 0 \\ 0 : & x < 0 \end{cases} \qquad (16)$$

and the water level $\gamma$ is chosen such that $\sum_{i=0}^{N-1} P_i = P_T$.

**Quadratic interpolation**

Using the three nearest quantized channel gains, the transmitter can perform quadratic interpolation to estimate other neighboring channel gains. Given $\alpha(Rk)$, $\alpha(R(k+1))$, and $\alpha(R(k+2))$, we want to find a quadratic function $q(x) = ax^2 + bx + c$, which satisfies Equation 17:

$$q(R(k+j)) = \alpha(R(k+j)), \quad j = 0,1,2. \qquad (17)$$

One formula for such a quadratic function is Lagrange's form of the interpolation polynomial given by Equation 18 (Abramovitz and Stegun, 1972):

$$q(x) = Rk \mathsf{L}_0(x) + R(k+1) \mathsf{L}_1(x) + R(k+2) \mathsf{L}_2(x) \quad (18)$$

where Lagrange basis functions are given by Equations 19–21:

$$\mathsf{L}_0(x) = \frac{1}{2R^2}(x - R(k+1))(x - R(k+2)), \qquad (19)$$

$$\mathsf{L}_1(x) = -\frac{1}{R^2}(x - R(k))(x - R(k+2)), \qquad (20)$$

$$\mathsf{L}_2(x) = \frac{1}{2R^2}(x - R(k))(x - R(k+1)). \qquad (21)$$

Thus, for $0 \le r \le 2R-1$ and $0 \le k \le K-3$, Equation 22 applies:

$$\hat{\alpha}(Rk + r) = q(Rk + r) \qquad (22)$$

where $q(\cdot)$ is defined in (17).

For the last two clusters, which may consist of fewer than $2R$ subcarriers, Lagrange basis functions are given by Equations 23–25:

$$\mathsf{L}_0(x) = \frac{(x - R(K-1))(x - N)}{R(N - R(K-2))}, \qquad (23)$$

$$\mathsf{L}_1(x) = -\frac{(x - R(K-2))(x - N)}{R(N - R(K-1))}, \qquad (24)$$

$$\mathsf{L}_2(x) = \frac{(x - R(K-2))(x - R(K-1))}{(N - R(K-1))(N - R(K-2))}. \qquad (25)$$

Similar to linear interpolation, the channel estimates obtained by quadratic interpolation can be used to find the water-filling power allocation. The resulting sum capacity from the water-filling solution based on either interpolation methods will depend on the subcarrier interval $R$ and the number of total feedback bits $B$. Higher-order channel interpolation can be obtained in a similar manner; however, the capacity gain is expected to be incremental.

**Bit allocation with channel interpolation**

Besides power allocation, the transmitter can use the estimate of the channel frequency response obtained by either linear or quadratic interpolation to allocate the transmitted bit. Let $C_b$ be the total bits transmitted for one OFDM symbol and hence Equation 26:

$$C_b = \sum_{i=0}^{N-1} c_i. \qquad (26)$$

Given $C_b$ and the BER, we would like to find the optimal bit allocation $\{c_i\}$ that minimizes the transmitted power. We apply the optimal bit allocation algorithm proposed by (Wong *et al.*, 1999). The algorithm assigns two bits at a time to a subcarrier, which requires the least amount of additional power. Thus, for every subcarrier, the transmitter computes the additional power by Equation 27:

$$\Delta P_i = f(c_i + 2) - f(c_i) \qquad (27)$$

Where Equation 28:

$$f(c_i) = \frac{\sigma_n^2 (2^{c_i} - 1)}{3[\hat{\alpha}(i)]^2} \left[ Q^{-1} \left( \frac{P_{e,i}}{4} \right) \right]^2 \qquad (28)$$

gives the transmitted power required to transmit $c_i$ bits and maintain the BER $P_{e,i}$, and is based on channel interpolation $\hat{\alpha}(i)$. The algorithm iterates until the sum of the transmitted bits over all subcarriers equals $C_b$ and subsequently obtains a set of transmitted power $P_i = f(c_i)$ for $0 \le i \le N-1$.

Assuming perfect channel information at the receiver, the system BER can be approximated as Equation 29:

$$\text{BER} \approx \frac{1}{N} \sum_{i=0}^{N-1} E\left[ 4Q\left( \sqrt{\frac{3P_t \left| H(i) \right|^2}{\sigma_n^2 2^{c_i-1}}} \right) \right], \qquad (29)$$

which depends on a number of feedback bits $B$, interpolation method, and cluster size $R$.

To illustrate the performance of the proposed methods, Monte Carlo simulation is used with 3,000 channel realizations. We also assume that the feedback channel is error- and delay-free.

The parameters used in the simulation are shown in Table 1.

## RESULTS AND DISCUSSION

Figure 1 shows the sum capacity of the on/off power allocation with different subcarrier clustering and numbers of clusters $K$ for $N = 128$, $P_T = 1$, and $\sigma_n^2 = 0.1$. As $K$ increases, the sum capacity increases. For the number of channel taps $M = 10$, the capacity almost achieves the maximum with $K = 32$ or 4 subcarriers per cluster. As the channel becomes less frequency-selective ($M = 5$), the number of subcarriers to achieve close to the maximum capacity is larger, for example, 16 in this figure. The capacity of the proposed on/off power allocation scheme was compared with the optimal water-filling capacity (Bingham, 1990; Tse and Visawanath, 2005) for $M = 10$, and approximately 10% performance degradation resulted when $K = 32$. However, the on/off scheme requires about 32 feedback bits, which is much lower than that required by the water-filling allocation. For the on/off power scheme, the optimal threshold $\mu$ was used and was found by a numerical search.

**Table 1** System parameters used in the simulation.

| Parameter | Notation | Value |
| --- | --- | --- |
| Number of total subcarriers | $N$ | 128 |
| Number of channel taps | $M$ | 3,5,6,8,10,12,20 |
| Noise variance | $\sigma_n^2$ | 0.1 |
| Number of clusters | $K$ | 1,2,4,8,16,32,64,128 |
| Number of subcarriers in one cluster | $R$ | 1,2,4,8,16,32,64,128 |
| Number of total feedback bits | $B$ | 32,64,128 |
| Number of total transmission bits | $C_b$ | 128 |
| Total transmission power | $P_T$ | 1 |

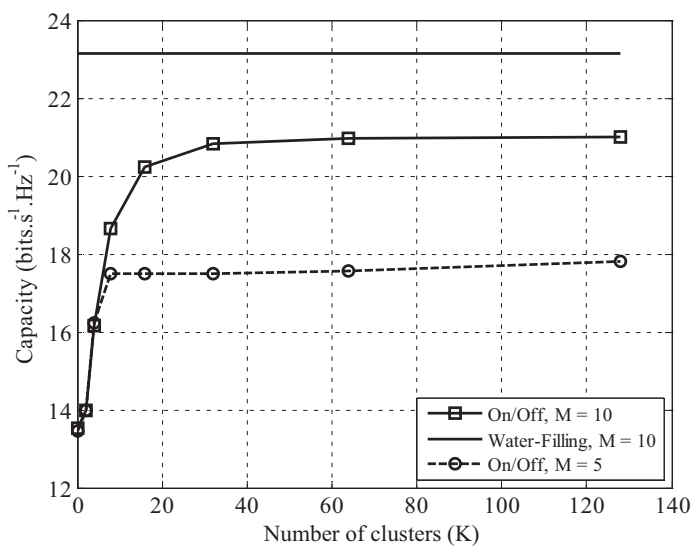

**Figure 1** Sum capacity of the proposed on/off power allocation with subcarrier clustering and optimal water-filling allocation with different numbers of clusters. ($M$ = number of channel taps; number of total subcarriers, $N$ = 128; total transmission power, $P_T = 1$ and noise variance $\sigma_n^2 = 0.1$.)

Figure 2 compares the sum capacity of water-filling power allocation with linear and quadratic channel interpolations. The total number of feedback bits is fixed at $B = 128$ bits. The quadratic channel interpolation performed better than the linear interpolation did, as was expected, and the performance gain from the linear interpolation was not significant. For both interpolation methods, the optimal value was $K = 15$. As $K$ increases, the number of channel gains that needs to be fed back increases and hence, each channel gain is quantized with fewer bits. Therefore, for very large $K$, the capacity is smaller.

The proposed quadratic interpolation scheme with optimal $K$ performs close to the water-filling capacity (within 3% difference) (Bingham, 1990; Tse and Visawanath, 2005). There was a uniform power allocation, which does not require any feedback, and a very large capacity gap between the uniform power and the proposed schemes (more than 30% performance gain).

Figure 3 shows the performance of the linear interpolation scheme for different $K$ values and also different numbers of total feedback bits $B$. The capacity as well as the optimal $K$ decreased with the available feedback bits. For $B = 128$, the optimal $K = 32$ while for $B = 64$, the optimal $K=8$. The result implies that for a very limited feedback rate, the cluster size $R$ should be large so that the number of channel gains to quantize is small. The capacity was degraded by as much as 6% when the feedback rate was reduced from 128 to 32 bits per update. Compared to the optimal water-filling power allocation (Bingham, 1990; Tse and Visawanath, 2005) in which perfect channel information is available at both the transmitter and the receiver, the proposed linear interpolation scheme performed within 2% of the optimum with optimal $K = 32$ for given feedback bits $B = 128$. The performance gap between the linear interpolation scheme and the water-filling allocation should narrow when more feedback is available. Furthermore, Figure 3 shows the performance of channel linear interpolation with a perfect set of squared channel gains (Athaudage and Jayalath, 2003), which in theory requires infinite feedback. As the number of clusters increased, the performance increased and approached the water-filling solution as expected. Finally, Figure 3 shows that setting the number of clusters to 32, or 4 subcarriers for one cluster in this example, achieves close to the optimum.

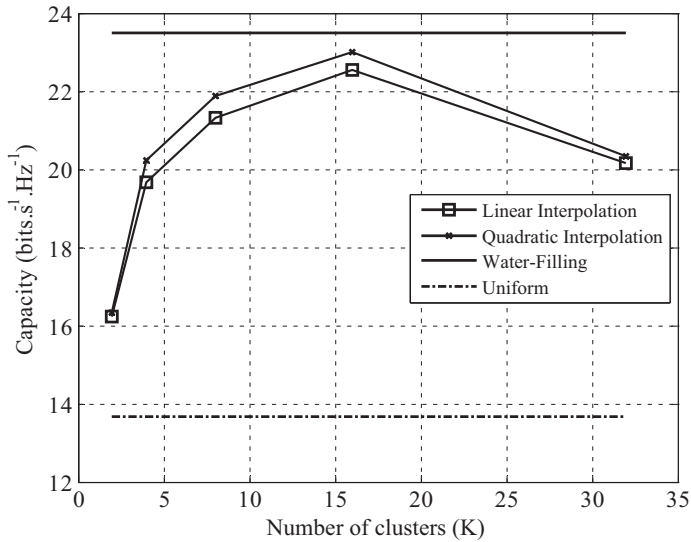

**Figure 2**  Sum capacity of linear and quadratic interpolation schemes for different number of clusters. (number of total subcarriers, $N = 128$; number of channel taps, $M = 10$; number of total feedback bits = 128; total transmission power, $P_T = 1$ and noise variance, $\sigma_n^2 = 0.1$.)

Figure 4 compares the sum capacity of all proposed schemes with the signal-to-noise ratio (SNR) and fixed number of feedback bits $B = 128$. The quadratic interpolation method outperformed the linear interpolation method and the on/off scheme for all SNR. All the proposed schemes performed close to the water-filling solution. In a low SNR regime, the proposed power allocation schemes, which in this example require 128 feedback bits, could outperform the uniform power allocation by 100%. In a high SNR regime, the effect of feedback was not as prominent.

Figure 5 compares the system BER from different schemes with the total SNR, which is the ratio between the total transmitted power over all subcarriers and the noise variance. For linear and quadratic interpolation, the BER is shown with various values of the cluster size $R$. To benchmark the proposed scheme, the BER is also shown with perfect channel information at the transmitter (Wong *et al.,* 1999). Selecting the right $R$ is critical. For $R = 8$, both interpolations with only one feedback bit per subcarrier performed close to an ideal system with perfect channel information at the transmitter. At a BER of $1 \times 10^{-3}$, selecting $R = 4$ required 1 dB more power than selecting $R = 8$. Generally, quadratic interpolation gave a small performance gain over linear interpolation.

Figure 6 examines how the cluster size affects a system BER of the proposed feedback schemes. Similar to the previous figure, the BER of an ideal system with perfect channel estimation (solid lines) is shown. In Figure 6, only the performance of linear interpolation is shown.

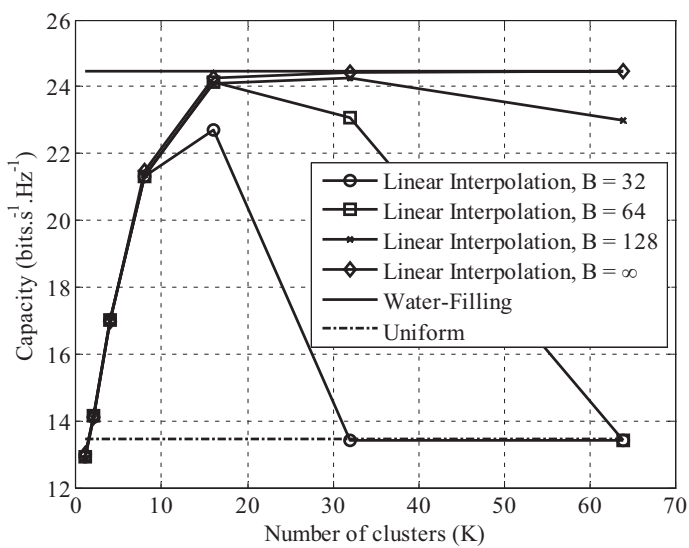

**Figure 3** Sum capacity of the linear interpolation scheme with the number of clusters and available feedback bits. (number of total subcarriers, $N = 128$; number of channel taps, $M = 10$; number of total feedback bits, $B = 128$; total transmission power, $P_T = 1$ and noise variance, $\sigma_n^2 = 0.1$.)

For different channel conditions ($M = 3, 12, 20$), selecting the optimal cluster size can be critical. Interpolating with optimal R gave the lowest bit error rate. For a channel with almost flat fading ($M = 3$), the optimal cluster size was large, ($R = 16$) due to the channel almost staying static for most of the subcarriers. As the channel became more frequency-selective (larger $M$), the optimal $R$ decreased. In Figure 6, for $M = 12$ and 20, the optimal $R$ is 8 and 4, respectively, and for those two cases, the optimal BER could be significantly smaller than the BER with arbitrary $R$.

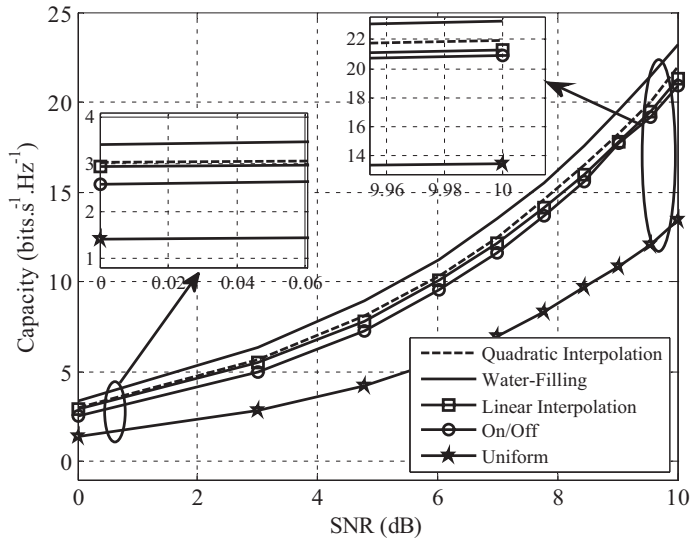

**Figure 4**  Capacity comparison of all proposed power allocation schemes with different signal-to-noise ratios (SNRs). (number of total subcarrieres, $N = 128$; number of channel taps, $M = 10$; total feedback bits, $B = 128$; and noise variance, $\sigma_n^2 = 0.1$.)

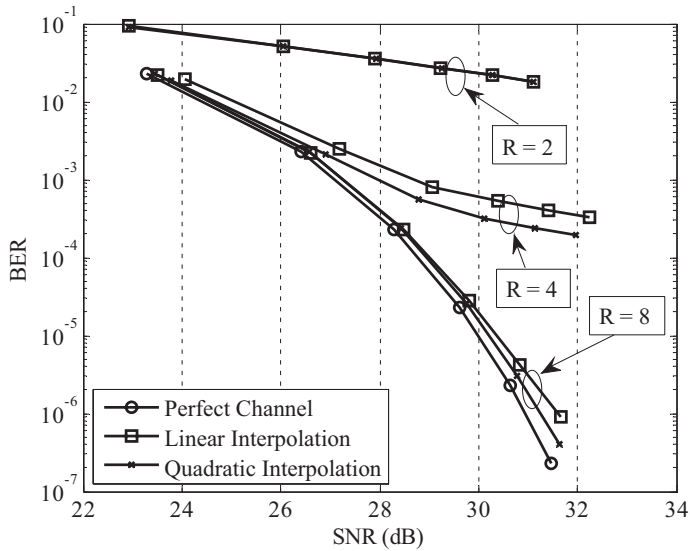

**Figure 5**  System bit error rate (BER) against signal-to-noise ratio with various numbers of subbcarriers in one cluster ($R$). (number of total subcarrieres, $N = 128$; number of channel taps, $M = 6$; total feedback bits, $B = 128$; total transmision bits, $C_b = 128$; and noise variance, $\sigma_n^2 = 0.1$.)

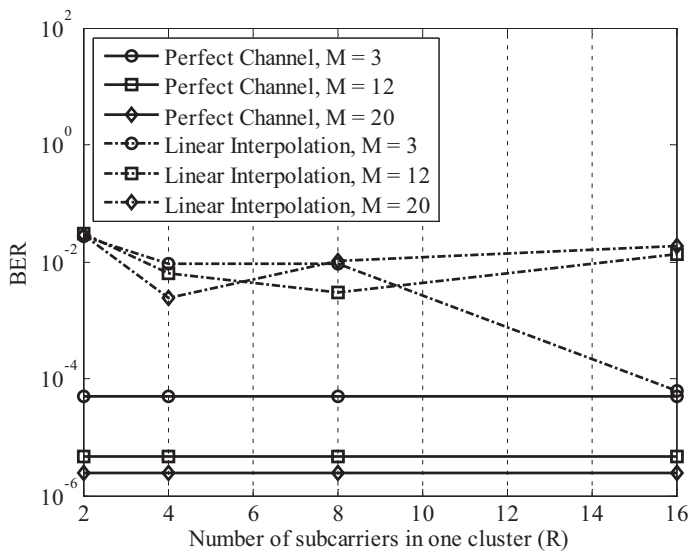

**Figure 6** Bit error rate (BER) with different numbers of subcarriers in one cluster (R) and different numbers of channel taps (M). (number of total subcarriers, $N = 128$; total feedback bits, $B = 64$; total transmision bits, $C_b = 128$; signal-to-noise ratio, SNR = 30 dB and noise variance $\sigma_n^2 = 0.1$.)

## CONCLUSION

Feedback schemes were proposed to allocate power levels and bits across subcarriers for a given feedback rate with the goal of reducing the feedback used to relay channel information to the transmitter while maintaining system performance. The numerical results showed that the water-filling allocation with quadratic channel interpolation performed best among the schemes and close to the optimum with unlimited feedback. For a given total transmission power and limited feedback rate, the proposed schemes could increase the spectral efficiency over a uniform-power transmission. For an actual implementation of adapting power levels in a wireless OFDM system, the on/off scheme with subcarrier clustering may be a more attractive choice due to much less computational complexity in the transmitter. The study also showed that selecting the optimal cluster size can give performance results close to those of an ideal system with perfect channel information. The performance of all proposed schemes depends on the cluster size, which is currently found by simulation. An analysis of the optimal cluster size is desirable and remains an open question.

## ACKNOWLEDGEMENTS

This work was supported by joint funding from the Thailand Commission on Higher Education, Thailand Research Fund, and Kasetsart University under grant MRG5580236, and the 2010 Telecommunications Research and Industrial Development Institute (TRIDI) scholarship. The material in this paper was presented in part at the international conference on information and communication technology for embedded systems (ICICTES), Samut Songkhram, Thailand, January 2013.